# DBAI

**T**ECHNICAL
**R** E P O R T

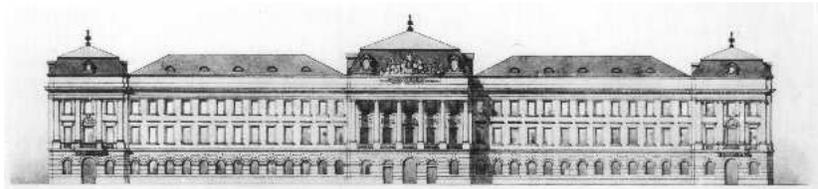

INSTITUT FÜR INFORMATIONSSYSTEME

ABTEILUNG DATENBANKEN UND ARTIFICIAL INTELLIGENCE

# Efficient generation of rotating workforce schedules

DBAI-TR-2000-35

**Nysret Musliu**    **Johannes Gärtner**    **Wolfgang Slany**

Institut für Informationssysteme
Abteilung Datenbanken und
Artificial Intelligence
Technische Universität Wien
Favoritenstr. 9
A-1040 Vienna, Austria
Tel:   +43-1-58801-18403
Fax:   +43-1-58801-18492
sekret@dbai.tuwien.ac.at
www.dbai.tuwien.ac.at

TECHNISCHE UNIVERSITÄT WIEN

# Efficient generation of rotating workforce schedules

**Nysret Musliu**,[1]    **Johannes Gärtner**,[2]    **Wolfgang Slany**,[3]

**Abstract.** Generating high-quality schedules for a rotating workforce is a critical task in all settings where a certain staffing level must be guaranteed beyond the capacity of single employees, such as for instance in industrial plants, hospitals, or airline companies. Results from ergonomics [2] indicate that rotating workforce schedules have a profound impact on the health and social life of employees as well as on their performance at work. Moreover, rotating workforce schedules must satisfy legal requirements and should also meet the objectives of the employing organization. We describe our solution to this problem. A basic design decision was to aim at quickly obtaining high-quality schedules for realistically sized problems while maintaining human control. The interaction between the decision maker and the algorithm therefore consists in four steps: (1) choosing a set of lengths of work blocks (a work block is a sequence of consecutive days of work shifts), (2) choosing a particular sequence of work and days-off blocks among those that have optimal weekend characteristics, (3) enumerating possible shift sequences for the chosen work blocks subject to shift change constraints and bounds on sequences of shifts, and (4) assignment of shift sequences to work blocks while fulfilling the staffing requirements. The combination of constraint satisfaction and problem-oriented intelligent backtracking algorithms in each of the four steps allows to find good solutions for real-world problems in acceptable time. Computational results from real-world problems and from benchmark examples found in the literature confirm the viability of our approach. The algorithms are now part of a commercial shift scheduling software package.

[1]Technische Universität Wien and Ximes Corp. mailto: musliu@dbai.tuwien.ac.at
[2]Ximes Corp. and Technische Universität Wien. mailto: gaertner@ximes.com
[3]Technische Universität Wien. mailto: wsi@dbai.tuwien.ac.at

**Acknowledgements**: This work was supported by FFF project No. **801160/5979** and the Austrian Science Fund Project No. **Z29-INF**.

# 1  Introduction

Workforce scheduling is the assignment of employees to shifts or days-off for a given period of time. There exist two main variants of this problem: rotating (or cyclic) workforce schedules and noncyclic workforce schedules. In a rotating workforce schedule—at least during the planning stage—all employees have the same basic schedule but start with different offsets. Therefore, while individual preferences of the employees cannot be taken into account, the aim is to find a schedule that is optimal for all employees on the average. In noncyclic workforce schedules individual preferences of employees can be taken into consideration and the aim is to achieve schedules that fulfill the preferences of most employees. In both variants of workforce schedules other constraints such as the minimum needed number of employees in each shift have to be satisfied. Both variants of the problem are NP-complete [8] and thus hard to solve in general, which is consistent with the prohibitively large search spaces and conflicting constraints usually encountered. For these reasons, unless it is absolutely required to find an optimal schedule, generation of good feasible schedules in a reasonable amount of time is very important. Because of the complexity of the problem and the relatively high number of constraints that must be satisfied, and, in case of soft-constraints, optimized, generating a schedule without the help of a computer in a short time is almost impossible even for small instances of the problem. Therefore, computerized workforce scheduling has been the subject of interest of researchers for more than 30 years. Tien and Kamiyama [11] give a good survey of algorithms used for workforce scheduling. Different approaches were used to solve problems of workforce scheduling. Examples for the use of exhaustive enumeration are [5] and [3]. Glover and McMillan [4] rely on integration of techniques from management sciences and artificial intelligence to solve general shift scheduling problems. Balakrishnan and Wong [1] solved a problem of rotating workforce scheduling by modeling it as a network flow problem. Smith and Bennett [10] combine constraint satisfaction and local improvement algorithms to develop schedules for anesthetists. Schaerf and Meisels [9] proposed general local search for employee timetabling problems.

In this paper we focus on the rotating workforce scheduling problem. The main contribution of this paper is to provide a new framework to solve the problem of rotating workforce scheduling, including efficient backtracking algorithms for each step of the framework. Constraint satisfaction is divided into four steps such that for each step the search space is reduced to make possible the use of backtracking algorithms. Computational results show that our approach is efficient for real-sized problems. The main characteristic of our approach is the possibility to generate high-quality schedules in short time interactively with the human decision maker.

The paper is organized as follows. In Section 2 we give a detailed definition of the problem that we consider. In Section 3 we present our new framework and our algorithms used in this framework. In Section 4 we discuss the computational results for two real-world problem and for problem instances taken from the literature. Section 5 concludes and describes work that remains to be done.

## 2 Definition of problem

In this section we describe the problem that we consider in this paper. The problem is a restricted case of a general workforce scheduling problem. General definitions of workforce scheduling problems can be found in [4, 9, 6]. The definition of the problem that we consider here is given below:

**INSTANCE:**

> Number of employees: $n$.
>
> Set $A$ of $m$ shifts (activities) : $a_1, a_2, \ldots, a_m$, where $a_m$ represents a day-off.
>
> $w$: length of schedule. The total length of a planning period is $n \times w$ because of the cyclicity of the schedules. Usually, $n \times w$ will be a multiple of 7, allowing to take weekends into consideration even when the schedule will be reused for more than one planning period.
>
> A cyclic schedule is represented by an $n \times w$ matrix $S \in A^{nw}$. Each element $s_{i,j}$ of matrix $S$ corresponds to one shift. Element $s_{i,j}$ shows in which shift employee $i$ works during day $j$ or whether the employee has free. In a cyclic schedule, the schedule of one employee consists in a sequence of all rows of the matrix $S$. The last element of a row is adjacent to the first element of the next row and the last element of the matrix is adjacent to its first element.
>
> Temporal requirements: $(m-1) \times w$ matrix $R$, where each element $r_{i,j}$ of matrix $R$ shows the required number of employees in shift $i$ during day $j$.
>
> Constraints:
>
> - Sequences of shifts allowed to be assigned to employees (the complement of forbidden sequences): Shift change $m \times m \times m$ matrix $C \in A^{(m^3)}$. If element $c_{i,j,k}$ of matrix $C$ is 1 then the sequence of shifts $(a_i, a_j, a_k)$ is allowed, otherwise not. Note that the algorithms we describe in Section 3 can easily be extended to allow longer allowed/forbidden sequences. Also note that Lau [8] could show that the problem is NP-complete even if we restrict forbidden sequences to length two.
> - Maximum and minimum of length of periods of successive shifts: Vectors $MAXS_m$, $MINS_m$, where each element shows maximum respectively minimum allowed length of periods of successive shifts.
> - Maximum and minimum length of work days blocks: $MAXW$, $MINW$

**PROBLEM:**

> Find as many non isomorphic cyclic schedules (assignments of shifts to employees) as possible that satisfy the requirement matrix, all constraints, and are optimal in terms of free weekends (weekends off).

The requirement matrix $R$ is satisfied if

$$\forall j \in \{1, 2, \ldots, w\} \ \forall k \in \{1, 2, \ldots, m-1\}$$
$$|\{i \in \{1, 2, \ldots, n\} / \ s_{i,j} = a_k\}| = r_{k,j}$$

The other constraints are satisfied if:

For the shift change matrix $C$:

$$\forall i \in \{1, \ldots, n\} \ \forall j \in \{1, \ldots, w\} \ \exists \ e, f, g \in \{1, \ldots, m\}$$
$$s_{i,j} = a_e \wedge next(s_{i,j}) = a_f \wedge next_2(s_{i,j}) = a_g \Rightarrow c_{e,f,g} = 1$$

where

$$next(s_{i,j}) = \begin{cases} s_{i,j+1} & \text{if } j < w \\ s_{i+1,1} & \text{if } j = w \text{ and } i < n \\ s_{1,1} & \text{otherwise} \end{cases}$$

and

$$next_k(s_{i,j}) = \underbrace{next(next(\ldots next(s_{i,j})\ldots))}_{k},$$

Maximum length of periods of successive shifts:

$$\forall k \in \{1, \ldots, m\} \ \forall i \in \{1, \ldots, n\} \ \forall j \in \{1, \ldots, w\}$$
$$(s_{i,j}, next(s_{i,j}), \ldots, next_{MAXS(k)}(s_{i,j})) \neq \underbrace{(a_k, \ldots, a_k)}_{MAXS(k)+1}$$

Minimum length of periods of successive shifts:

$$\forall k \in \{1, \ldots, m\} \ \forall i \in \{1, \ldots, n\} \ \forall j \in \{1, \ldots, w\}$$
$$\neg(s_{i,j} \neq a_k \wedge next(s_{i,j}) = a_k \wedge (\bigvee_{b \in \{2, \ldots, MINS(k)\}} next_b(s_{i,j}) \neq a_k))$$

Maximum length of work blocks:

$$\forall i \in \{1, \ldots, n\} \ \forall j \in \{1, \ldots, w\}$$
$$(s_{i,j} = a_m \vee next(s_{i,j}) = a_m \vee \ldots \vee next_{MAXW}(s_{i,j}) = a_m)$$

Minimum length of work blocks:

$$\forall i \in \{1, \ldots, n\} \ \forall j \in \{1, \ldots, w\}$$
$$\neg(s_{i,j} = a_m \wedge next(s_{i,j}) \neq a_m \wedge (\bigvee_{b \in \{2, \ldots, MINW\}} next_b(s_{i,j}) = a_m))$$

The rationale behind trying to obtain more than one schedule will be made clear in Section 3.

Optimality of free weekends is in most of the times in conflict with the solutions selected for work blocks.

## 3 Four step framework

Tien and Kamiyama [11] proposed a five stage framework for workforce scheduling algorithms. This framework consist of these stages: determination of temporal manpower requirements, total manpower requirement, recreation blocks, recreation/work schedule and assignment of shifts (shift schedule). The first two stages can be seen as an allocation problem and the last three stages are days-off scheduling and assignment of shifts. All stages are related with each other and can be solved sequentially, but there exist also algorithms which solve two or more stages simultaneously.

In our problem formulation we assume that the temporal requirements and total requirements are already given. Temporal requirements are given through the requirement matrix and determine the number of employees needed during each day in each shift. Total requirements are represented in our problem through the number of employees $n$. We propose a new framework for solving the problem of assigning days-off and shifts to the employees. This framework consist of the following four steps:

1. choosing a set of lengths of work blocks (a work block is a sequence of consecutive days of work shifts),

2. choosing a particular sequence of work and days-off blocks among those that have optimal weekend characteristics,

3. enumerating possible shift sequences for the chosen work blocks subject to shift change constraints and bounds on sequences of shifts, and

4. assignment of shift sequences to work blocks while fulfilling the staffing requirements.

First we give our motivation for using this framework. Our approach is focused on the interaction with the decision maker. Thus, the process of generating schedules is only half automatic. When our system generates possible candidate sets of lengths of work blocks in step 1 the decision maker will select one of the solutions that best reflects his preferences. This way we satisfy two goals: On the one hand, an additional soft constraint concerning the lengths of work blocks can be taken into account through this interaction, and on the other hand the search space for step 2 is significantly reduced. Thus we will be able to solve step 2 much more effectively. In step 2 our main concern is to find a best solution for weekends off. The user selection in step 1 can impact features of weekends off versus length of work blocks since these two constraints are the ones that in practice most often are in conflict. The decision maker can decide if he wishes optimal length of work blocks or better features for weekends off. With step 3 we satisfy two more goals. First, because of the shift change constraints and the bounds on the number of successive shifts in a sequence, each work block has only few legal shift sequences (terms) and thus in step 4 backtracking algorithms will find very fast assignments of terms to the work blocks such that the requirements are fulfilled (if shift change constraints with days-off exist, their satisfaction is checked at this stage). Second, a new soft constraint is introduced. Indeed, as we generate a bunch of shift plans, they will contain different terms. The user has then the possibility to eliminate some undesired terms, thus eliminating solutions that contain these terms. Terms can have an impact on fatigue and sleepiness of the employees and as such are very important when high-quality plans are sought.

Table 1: A possible schedule with work blocks in the order (46546555)

| Employee/day | Mon | Tue | Wen | Thu | Fri | Sat | Sun |
|---|---|---|---|---|---|---|---|
| 1 | D | D | A | A |   |   |   |
| 2 | A | A | A | N | N | N |   |
| 3 |   | D | D | N | N | N |   |
| 4 |   | A | A | A | A |   |   |
| 5 |   |   | D | D | D | D | D |
| 6 | D |   |   | D | D | D | N |
| 7 | N |   |   |   | A | A | N |
| 8 | N | N |   |   |   | A | A |
| 9 | A | N | N |   |   |   |   |

## 3.1 Determination of lengths of work blocks

A work block is a sequence of work days between two days-off. An employee has a work day during day $j$ if he/she is assigned a shift different from the days-off shift $a_m$. In this step the feature of work blocks in which we are interested is only its length. Other features of work blocks (e.g., shifts of which the work block is made of, begin and end of block, etc.) are not known at this time. Because the schedule is cyclic each of the employees has the same schedule and thus the same work blocks during the whole planning period.

**Example:** The week schedule for 9 employees given in Table 1 consists of two work blocks of length 6, four work blocks of length 5 and two of length 4 in the order (4 6 5 4 6 5 5 5). By rearranging the order of the blocks, other schedules can be constructed, for example the schedule with the order of work blocks (5 5 6 5 4 4 5 6). We will represent schedules with the same work blocks but different order of work blocks through unique solutions called class solution where the blocks are given in decreasing order. The class solution of above example thus will be {6 6 5 5 5 5 4 4}.

It is clear that even for small instances of problems there exist many class solutions. Our main concern in this step is to generate all possible class solution or as many as possible for large instances of the problem.

One class solution is nothing else than an integer partition of the sum of all working days that one employee has during the whole planning period. To find all possible class solutions in this step we have to deal with the following two problems:

- Generation of restricted partitions, and

- Elimination of those partitions for which no schedule can be created.

Because the elements of a partition represent lengths of work blocks and because constraints about maximum and minimum length of such work blocks exist, not all partitions must be generated. The maximum and minimum lengths of days-off blocks also impact the maximum and minimum allowed number of elements in one partition, since between two work blocks there always is a days-off block, or recreation block. In summary, partitions that fulfill the following criteria have to be generated:

- Maximum and minimum value of elements in a partition. These two parameters are respectively maximum and minimum allowed length of work blocks,

- Minimum number of elements in a partition:

$$MINB = \left\lceil \frac{DaysOffSum}{MAXS(m)} \right\rceil$$

$DaysOffSum$: Sum of all days-off that one employee has during the whole planning period.

- Maximum number of elements in a partition:

$$MAXB = \left\lfloor \frac{DaysOffSum}{MINS(m)} \right\rfloor$$

The set of partitions which fulfill above criteria is a subset of the set of all possible partitions. One can first generate the full set of all possible partitions and then eliminate those that do not fulfill the constraints given by the criteria. However, this approach is inefficient for large instances of the problem. Our idea was to use restrictions for pruning while the partitions are generated. We implemented a procedure based in this idea for the generation of restricted partitions. Pseudo code is given below. The set $P$ contains elements of a partition of $N$.

Initialize $N$, $MAXB$, $MINB$, $MAXW$, $MINW$
'Value of arguments for first procedure call
$Pos = 1$, $MaxValue = MAXW$

'Recursive procedure
RestrictedPartitions($Pos$, $MaxValue$)

    $i = MINW$
    Do While ($i <= MaxValue \land \neg PartitionIsCompleted$)

        Add to the set $P$ element $i$
        $PSum =$ Sum of elements in a set $P$

        If ($PSum = N \land Pos \geq MINB$) Then

        $PartitionIsCompleted =$ true
        Store partition (set $P$)
     'Pruning
     ElseIf $(Pos < MAXB \wedge PSum \leq N - MINW)$ Then
        'Recursive call
        RestrictedPartitions($Pos + 1, i$)
     EndIf
     $i = i + 1$
     Remove last element from set $P$
  Loop

End

  Not all restricted partitions can produce a legal schedule that fulfills the requirements of work force per day (in this step we test if we have the desired number of employees for a whole day, not for each shift). As we want to have only class solutions that will bring us to a legal shift plan, we eliminate all restricted partitions that cannot fulfill the work force per day requirements. A restricted partition will be legal if at least one distribution of days-off exists that fulfills the work force per day requirements. In the worst case all distributions of days-off have to be tested if we want to be sure that a restricted partition has no legal days-off distribution. One can first generate all days-off distributions and then test each permutation of restricted partitions if at least one satisfying days-off distribution can be found. This approach is rather ineffective when all class solution have to be generated because many of them will not have legal days-off distribution and thus the process of testing takes too long for large instances of typical problems. We implemented a backtracking algorithms for testing the restricted partitions. Additionally, as we want to obtain the first class solutions as early as possible, we implemented a three stages time restricted test. In this manner we will not lose time with restricted partitions which do not have legal days-off distribution in the beginning of test. The algorithm for testing the restricted partitions is given below (without the time restriction).

INPUT: Restricted partition, possible days-off blocks.
Initialize vectors $W(NumberOfUniqueWorkBlocks)$ and
$F(NumberOfUniqueDaysOffBlocks)$ with unique work blocks, respectively with unique days-off blocks (for example unique work blocks of class solution {5 5 4 4 3} are blocks 5, 4 and 3).

'$i$ represents one work or days-off block. It takes values from 1 to $numberOfWorkBlocks * 2$ (after each work block comes a days-off block and our aim is to find the first schedule that fulfills the requirements per day). For the first procedure call $i = 1$
'Recursive procedure
PartitionTest($i$)

$k = 1$
If $i$ is odd

 Do While($k \leq NumberOfUniqueWorkBlocks$)

  Assign block $i$ with work block $W(k)$

  If $i = LastBlock - 1$ then

   $Req=$ for each day the number of employees does not get larger than the requirement and the number of work blocks of type $W(k)$ does not get larger than the number of blocks of type $W(k)$ in the class solution
   'Pruning
   If $Req =$true then
    PartitionTest($i + 1$)
   Endif

  Else

   'Only partial schedule until block $i$ is tested
   $Req =$ requirements for number of employees during each day are not overfilled and number of work blocks of type $W(k)$ does not get larger than the number of blocks of type $W(k)$ in the class solution

   If $Req =$true then
    PartitionTest($i + 1$)
   Endif

  Endif
  $k = k + 1$

 Loop

Else

 Do While($k \leq NumberOfUniqueDaysOffBlocks$)

  Assign block $i$ with days-off block $F(k)$

  If $i = lastblock$ then

   $SumTest =$Test if sum of all days-off is as required
   If $SumTest =$true then
    Class solution has at least one days-off distribution
    Interrupt test
   Endif

```
            Else
                'Only partial schedule until block i is tested
                FreeTest =Test if not more employees than required have
                free (test is done for each day)
                'Pruning
                If FreeTest =true then
                    PartitionTest(i + 1)
                Endif
            Endif
            k = k + 1
        Loop
    Endif
End
```

## 3.2 Determination of distribution of work blocks and days-off blocks that have optimal weekend characteristics

Once the class solution is known different shift plans can be produced subject to the order of work blocks and to the distribution of days-off. For each order of blocks of the class solution there may exist many distributions of days-off. We introduce here a new soft constraint. This constraint concerns weekends off. Our goal here is to find for each order of work blocks the best solution (or more solutions if they are not dominated) for weekends off. The goal is to maximize the number of weekends off, maximize the number of long weekends off (the weekend plus Monday or Friday is free) and to find solution that have a "better" distribution of weekends off. Distribution of weekends off will be evaluated with the following method: Every time two weekends off appear directly after each other the distribution gets a negative point. One distribution of weekends off is better than another one if it has less negative points. Priority is given to the number of weekends off followed by the distribution of weekends off and finally the number of long weekends off is considered only if the others are equal. Possible candidates are all permutations of the work blocks found in a class solution. Each permutation may or may not have days-off distributions. If the permutation has at least one days-off distribution our goal is to find the best solutions for weekends off. The best solutions are those that cannot be dominated by another solution. We say that solution $Solut_1$ dominates solution $Solut_2$ in the following cases:

- $Solut_1$ has the same number of weekends off as $Solut_2$, the evaluation of the weekends distribution of $Solut_1$ is equal to the one of $Solut2$, and $Solut_1$ has more long weekends off than $Solut_2$

- $Solut_1$ has same number of weekends off as $Solut_2$ and the evaluation of the weekends distribution of $Solut_1$ is better than the one of $Solut2$

- $Solut_1$ has more weekends off than $Solut_2$

Two comments have to be made here. First, because some of the permutations of the class solutions may not have any days-off distribution, we use time restrictions for finding days-off distributions. In other words, if the first days-off distribution is not found in a predetermined time, the next permutation is tested. Second, for large instances of problems, too many days-off distributions may exist and this may impede the search for the best solution. Interrupting the test can be done manually depending on the size of the problem.

For large instances of the problem it is impossible to generate all permutations of class solutions and for each permutation the best days-off distributions. In these cases our main concern is to enumerate as many solutions as possible which have the best day-off distribution and can be found in a predetermined time. Found solutions are sorted based in weekends attributes such that the user can decide easier which distribution of days-off and work days he wants to continue with. The user may select one of the solutions solely based in weekends, but sometimes the order of work blocks may also decide. One can prefer for example the order of work blocks (7 6 3 7 6 3 7 6) to the order (7 7 7 6 6 6 3 3).

For finding legal days-off distributions for each permutation of a class solution we use a backtracking procedure similar to the one for testing the restricted partitions in step 1 except that the distribution of work blocks is now fixed. After the days-off distributions for a given order of work blocks are found, selecting the best solutions based on weekends is a comparatively trivial task and takes not very long.

Selected solutions in step 2 have a fixed distribution of work blocks and days-off blocks, and in the final step the assignment of shifts to the work blocks has to be done.

## 3.3 Generating allowed shift sequences for each work block

In step 2 work and days-off blocks have been fixed. It remains to assign shifts to the employees. We again use a backtracking algorithm, but to make this algorithm more efficient we introduce another interaction step. The basic idea of this step is this: For each work block construct the possible sequences of shifts subject to the shift change constraints and the upper and lower bounds on the length of sequences of successive same shifts. Because of these constraints, the number of such sequences (we will call them terms) is not too large and thus backtracking algorithms will be much more efficient compared to classical backtracking algorithms where for each position of work blocks all shift possibilities would have to be tried and the test for shift change constraints would have to be done in a much more time-consuming manner, thus resulting in a much slower search for solutions.

**Example**: Suppose that the solution selected by the user in step 2 has the distribution of work blocks (6 4 4 6 5 4 5).
Shifts: Day (D), Afternoon (A) and Night (N)
Forbidden shift changes: (N D), (N A), (A D)
Maximum and minimum lengths of successive shifts: D: 2-6, A: 2-5, N: 2-4

Our task is to construct legal terms for work blocks of length 6, 5, and 4.

For work block of legth 6 the following terms exist:
DDDDDD, DDDDAA, DDDDNN, DDDAAA, DDDNNN, DDAAAA, DDNNNN, DDAANN, AAAANN, AAANNN, AANNNN

Block of length 5:
DDDDD, DDDAA, DDDNN, DDAAA, DDNNN, AAAAA, AAANN, AANNN

Block of length 4:
DDDD, DDAA, DDNN, AAAA, AANN, NNNN

This approach is very appropriate when the number of shifts is not too large. When the number of shifts is large we group shifts with similar characteristics in so called shift types. For example if there exists a separate day shift for Saturday which begins later than the normal day shift, these two shifts can be grouped together. Such grouping of similar shifts in shift types allows us to have a smaller number of terms per work blocks and therefore reduces the overall search space. To the end a transformation from shift types to the substituted shifts has to be done. A similar approach has been applied by Weil and Heus [12]. They group different days-off shifts in one shift type and thus reduce the search space. Different days-off shifts can be grouped in one shift only if they are interchangeable (the substitution has no impact in constraints or evaluation).

The process of constructing the terms takes not long most of the time given that the length of work blocks usually is less than 9 and some basic shift change constraints always exist because of legal working time restrictions.

## 3.4 Assignment of shift sequences to work blocks

Once we know the terms we can use a backtracking algorithm to find legal solutions that satisfy the requirements in every shift during every day. The size of the search space that should be searched with this algorithm is:

$$\Pi_{i=1}^{b} N_t(i)$$

where $b$ is the number of work blocks and $N_t(i)$ is the number of legal terms for block $i$.

If we would not use terms the search space would be of size:

$$(m-1)^{\text{sum of all work days}}$$

Of course in this latter case we would have more constraints, for instance the shift change constraints, but the corresponding algorithm would be much slower because the constraints are tested not only one time as when we construct the terms.

Pseudo code for the backtracking algorithm based on terms is given below. Let us observe that the terms tests for shift change constraints is done without consideration of shift $a_m$ (days-off). If there exist shift changes constraints that include days-off, then test of the solution has to be done later for these sequences.

INPUT: distribution of work and days-off blocks

Generate all legal shift sequences for each work block
'Value of argument for first call of procedure ShiftAssignment
$i = 1$

'Recursive procedure
ShiftAssignment($i$)

    $j$ = Number of shift sequences of block $i$
    $k = 1$
    Do While ($k \leq j$)

        Assign block $i$ with sequence number $k$
        If $i = lastblock$ then

            $Req$ = Test if requirements are fulfilled and shift change constraints are not violated (in this stage we test for forbidden shift sequences that include days-off)
            If $Req$ = true then
                Store the schedule
            Endif

        Else

            'Only partial schedule until block i is tested
            $PTest$=Test each shift if not more than needed employees are assigned to it
            'Pruning ...
            If $Ptest$ =true then
                ShiftAssignment($i + 1$)
            Endif

        Endif
        $k = k + 1$

    Loop

End

There exist rare cases when even if there exists a work and days-off distribution no assignment of shifts can be found that fulfills the temporal requirements for every shift in every day because of shift change constraints. In these cases constraints about minimum and maximum length of periods of successive shifts must be relaxed to obtain solutions.

## 4   Computational results

In this section we report on computational results obtained with our approach. We implemented our four step framework in a software package called First Class Schedule (FCS) which is part of a shift scheduling package called Shift-Plan-Assistant (SPA) of XIMES[1] Corp. All our results in this section have been obtained on an Intel P2 330 Mhz. Our first two examples are taken from a real-world sized problems and are typical for the kind of problems for which FCS was designed. After that, we give our results for three benchmark examples from the literature and compare them with results from a paper of Balakrishnan and Wong [1] . They solved problems of rotating workforce scheduling through the modeling in a network flow problem. Their algorithms were implemented in Fortran on an IBM 3081 computer.

**Problem 1**: An organization operates in one 8 hours shift: Day shift (D). ¿From Monday to Saturday 4 employees are needed, whereas on Sunday no employees are needed. These requirements are fulfilled with 5 employees which work on average 38,4 hours per week. A rotating week schedule has to be constructed which fulfills the following constraints:

1. Length of periods of successive shifts should be: D: 2-6

2. Length of work blocks should be between 2 and 6 days and length of days-off blocks should be between 1 and 4

3. Features for weekends off should be as good as possible

We note that days-off blocks of length 1 are not preferred, but otherwise no class solution for this problem would exist.

Using FCS in step 1, all class solutions are generated in 4.5 seconds: {6 4 4 3 3 2 2}, {6 6 3 3 2 2 2}, {6 6 4 3 3 2}, {6 6 4 4 2 2}, {6 6 6 2 2 2}, {6 6 6 3 3}. We select the class solution {6 6 4 4 2 2} to proceed in the next step. In step 2 FCS generates 6 solutions after 0.6 seconds. Each of them has one long weekend off. We select a solution with the distribution of work blocks (6 4 4 6 2 2) to proceed in the next steps. Step 3 and 4 are solved automatically. We obtain the first and only existing schedule after 0.02 seconds. This solution is shown in Table 2.

The quality of this schedule stems from the fact that there are at most 8 consecutive work days with only a single day-off in between them. This constraint is very important when single days-off are allowed. This example showed a small instance of a problem with only one shift; nevertheless, even for a such instances it is relatively difficult to find high quality solutions subject to this constraint.

---

[1] http://www.ximes.com/

Table 2: First Class Schedule solution for problem 1

| Employee/day | Mon | Tue | Wed | Thu | Fri | Sat | Sun |
|---|---|---|---|---|---|---|---|
| 1 | D | D | D | D | D | D |   |
| 2 |   |   | D | D | D | D |   |
| 3 | D | D | D | D |   |   |   |
| 4 | D | D | D | D | D | D |   |
| 5 | D | D |   |   | D | D |   |

Let us note here that the same schedule can be applied for a multiple of 5 employees (the duties are also multiplied) if the employees are grouped in teams. For example if there are 30 employees they can be grouped in 5 teams. Each of them will have 6 employees.

**Problem 2**: An organization operates in three 8 hours shifts: Day shift (D), Afternoon shift (A), and Night shift (N). From Monday to Friday three employees are needed during each shift, whereas on Saturday and Sunday two employees suffice. These requirements are fulfilled with 12 employees which work on average 38 hours per week. A rotating week schedule has to be constructed which fulfills the following constraints:

1. Sequences of shifts not allowed to be assigned to employees are:
   (A D), (N D), (N A)

2. Length of periods of successive shifts should be: D: 2-7, A: 2-6, N: 2-5

3. Length of work blocks should be between 4 and 7 days and length of days-off blocks should be between 2 and 4

4. Features for weekends off should be as good as possible

Using FCS in step 1, the first class solution is generated after 0.07 seconds and we interrupt the process of generation of class solutions after 1.6 seconds when already 7 class solutions have been generated out of many others: {6 6 5 5 5 5 5 5 5 5}, {6 6 6 5 5 5 5 5 5 4}, {7 6 5 5 5 5 5 5 5 4}, {7 6 6 5 5 5 5 5 4 4}, {7 7 5 5 5 4 4 4 4 4 4}, {7 7 7 6 5 5 5 5 5 5}, and {7 7 7 6 6 5 5 5 4}. The first solution has the highest number of most optimal blocks, namely those with length 5, but entails weak features for weekends off. For this reason we select the class solution {7 7 7 6 5 5 5 5 5 5} to proceed in the next step. In step 2 the optimal solution for the distribution of blocks (7 7 7 6 5 5 5 5 5 5), with 6 weekends off from which 3 are long, is found in less than 2 seconds. The first 11 solutions are generated in 11 seconds, where we have one solution with 6 weekends off, 4 of which are long, and a distribution of weekends off that is acceptable. This solution has the order of work blocks as follows: (7 7 6 5 5 5 7 5 5 5). We select this solution to proceed in the next steps. Step 3 and 4 are solved automatically. We obtain a first schedule which is given in Table 3 after 0.17 seconds and the first 50 schedules after 4 seconds. The decision maker can eliminate some undesired terms. Besides these solutions there exist also a large amount of other solutions which differ in terms with each other. If a better distribution of weekends off would have been sought, this could have been found through another class solution, for example {7 7 7 7 7 7 5 5 5} found in step 1 after 16 seconds, at the cost of longer work sequences.

Table 3: First Class Schedule solution for problem 2

| Employee/day | Mon | Tue | Wed | Thu | Fri | Sat | Sun |
|---|---|---|---|---|---|---|---|
| 1  | D | D | D |   |   | A | A |
| 2  | A | A | A | A |   |   |   |
| 3  | D | D | D | D | D |   |   |
| 4  |   |   | A | A | A | N | N |
| 5  |   |   |   |   | N | N | N |
| 6  | N | N |   |   | A | A | A |
| 7  | A | A | N | N |   |   |   |
| 8  | A | A | A | A | A |   |   |
| 9  | N | N | N | N | N |   |   |
| 10 | N | N | N | N | N |   |   |
| 11 |   | D | D | D | D | D | D |
| 12 | D |   |   | D | D | D | D |

**Problem 3**: The first problem from literature for which we discuss computational results for First Class Schedule is a problem solved by Butler [3] for the Edmonton police department in Alberta, Canada. Properties of this problem are:

Number of employees: 9
Shifts: 1 (Day), 2 (Evening), 3 (Night)

Temporal requirements:

$$R_{3,7} = \begin{pmatrix} 2 & 2 & 2 & 2 & 2 & 2 & 2 \\ 2 & 2 & 2 & 3 & 3 & 3 & 2 \\ 2 & 2 & 2 & 2 & 2 & 2 & 2 \end{pmatrix}$$

Constraints:

- Length of work periods should be between 4 and 7 days

- Only shift 1 can precede the rest period preceding a shift 3 work period

- Before and after weekends off, only shift 3 or shift 2 work periods are allowed

- At least two consecutive days must be assigned to the same shift

Table 4: Solution of Balakrishnan and Wong [1] for the problem from [3]

| Employee/day | Mon | Tue | Wed | Thu | Fri | Sat | Sun |
|---|---|---|---|---|---|---|---|
| 1 | A | A | A | A | A | A |   |
| 2 |   | N | N | N | N | N |   |
| 3 |   |   | A | A | A | A | A |
| 4 |   |   | D | D | D | D | D |
| 5 | D | D |   |   | N | N | N |
| 6 | N |   |   | A | A | A | A |
| 7 | A | A |   |   | D | D | D |
| 8 | D | D | D | D |   |   | N |
| 9 | N | N | N | N |   |   |   |

- No more than two 7-day work periods are allowed and these work periods should not be consecutive

Balakrishnan and Wong [1] solve this problem using a network model and they needed 73.54 seconds to identify an optimal solution of the problem. This solution is given in Table 4. We use D for 1, A for shift 2, N for shift 3, and if the element of the matrix is empty the employee has free.

Before we give our computational results some observations should be made. First constraint two and three cannot be represented in our framework. Let us note here that in all three examples given, we cannot model the problem exactly (the same was true for Balakrishnan and Wong's [1] approach to the original problems), which is to a high degree due to the different legal requirements found in the U.S./Canadian versus those found in the European context, but we tried to mimic the constraints as closely as possible or to replace them by similar constraints that appeared more meaningful in the European context. Having said this, let us proceed as follows: The other constraints can be applied in our model and are left like in the original problem. As mentioned, we include additional constraints about maximum length of successive shifts and minimum and maximum length of days-off blocks. In summary, additional constraint used for First Class Schedule are:

- Not allowed shift changes: (N D), (N A), (A D)

- Length of days-off periods should be between 2 and 4

- Vector $MAXS_3 = (7, 6, 4)$

In our model we first generate class solutions. Class solutions that exist for the given problem and given constraints are:
$\{6\,6\,6\,6\,6\,5\,5\,5\}$, $\{7\,6\,6\,6\,6\,5\,5\,4\}$, $\{7\,7\,7\,7\,7\,5\,5\}$, $\{6\,6\,6\,6\,6\,6\,5\,4\}$, $\{7\,6\,6\,6\,5\,5\,5\,5\}$, $\{7\,6\,6\,6\,6\,6\,4\,4\}$, $\{7\,7\,6\,6\,5\,5\,5\,4\}$, $\{7\,7\,6\,6\,6\,5\,4\,4\}$, $\{7\,7\,7\,4\,4\,4\,4\,4\}$, $\{7\,7\,7\,5\,5\,5\,5\,4\}$, $\{7\,7\,7\,6\,5\,5\,4\,4\}$,

Table 5: First Class Schedule solution for the problem from [3]

| Employee/day | Mon | Tue | Wed | Thu | Fri | Sat | Sun |
|---|---|---|---|---|---|---|---|
| 1 | D | D | D |   |   | D | D |
| 2 | D | N | N |   |   | A | A |
| 3 | A | N | N | N | N |   |   |
| 4 | A | A | A | A | A |   |   |
| 5 |   | D | D | N | N | N |   |
| 6 |   | A | A | A | A | A | A |
| 7 |   |   |   | D | D | N | N |
| 8 | N |   |   | A | A | A | N |
| 9 | N |   |   | D | D | D | D |

{7 7 7 6 6 4 4 4}, {7 7 7 6 6 6 6}, {7 7 7 7 5 4 4 4}, {7 7 7 7 6 6 5}, {7 7 7 7 7 6 4}, {7 7 6 5 5 5 5 5}.

The first solution is generated in 0.14 seconds and all solutions in 4.38 seconds. We select in this step the class solution with the highest number of optimal blocks: {7 7 6 5 5 5 5 5}

In step 2 the distributions of work and days-off periods that gives best results for weekends is found. We select the best solution offered in this step from first class schedule that has this order of work blocks (7 5 7 5 5 6 5 5 ). The computations of the system took 0.73 seconds.

Step 3 and 4 are solved automatically. The first solution given in Table 5 is generated after 0.39 seconds and the first 50 solutions in 4.29 seconds.

There exist also many other solutions that differ only in terms that they contain. Undesired solutions can then be eliminated through the elimination of unwanted terms.

**Problem 4** (Laporte et al. [7]): There exist three non overlapping shifts D, A, and N, 9 employees, and requirements are 2 employees in each shift and every day. A week schedule has to be constructed that fulfills these constraints:

1. Rest periods should be at least two days-off

2. Work periods must be between 2 and 7 days long if work is done in shift D or A and between 4 and 7 if work is done in shift N

3. Shift changes can occur only after a day-off

4. Schedules should contain as many weekends as possible

5. Weekends off should be distributed throughout the schedule as evenly as possible

6. Long (short periods) should be followed by long (short) rest periods

Table 6: Solution of Balakrishnan and Wong [1] of problem from [7]

| Employee/day | Mon | Tue | Wed | Thu | Fri | Sat | Sun |
|---|---|---|---|---|---|---|---|
| 1 | A | A | A | A |   |   | D |
| 2 | D | D | D | D |   |   |   |
| 3 |   | D | D | D | D | D |   |
| 4 |   |   | A | A | A | A | A |
| 5 |   |   | N | N | N | N | N |
| 6 | N | N |   |   | A | A | A |
| 7 | A | A |   |   | D | D | D |
| 8 | D |   |   | N | N | N | N |
| 9 | N | N | N |   |   |   |   |

7. Work periods of 7 days are preferred in shift N

Balakrishnan and Wong [1] need 310.84 seconds to obtain the first optimal solution. The solution is given in Table 6. The authors report also about another solution with three weekends off found with another structure of costs for weekends off.

In FCS constraint 1 is straightforward. Constraint 2 can be approximated if we take the minimum of work blocks to be 4. Constraint 3 can also be modeled if we take the minimum length of successive shifts to be 4. For maximum length of successive shifts we take 7 for each shift. Constraints 4 and 5 are incorporated in step 2, constraint 6 cannot be modeled, and constraint 7 is modeled by selecting appropriate terms in step 3.

With these given parameters for this problem there exist 23 class solutions which are generated in 5 seconds. For each class solution there exist at least one distribution of days-off, but it could be that no assignment of shifts to the work blocks exist because the range of blocks with successive shifts are too narrow in this case. Because in this problem the range of lengths of blocks of successive shifts is from 4 to 7 for many class solution no assignment of shifts can be found. Class solution {7 7 6 5 5 4 4 4} gives solutions with three free weekends, but they are after each other. Class solution {7 7 7 7 5 5 4} gives better distribution of weekends. If we select this class solution in step 1, our system will generate 5 solutions in step 2 in 1.69 seconds. We selected a solution with the order of work blocks to be (7 7 4 7 5 7 5). Step 3 and 4 are solved simultaneously and the first solution was arrived at after 0.08 seconds. 18 nonisomorphic solutions were found after 0.5 seconds. One of the solutions is shown in Table 7.

With class solution {7 7 7 7 7 7} the same distribution of weekends off can be found as in [7].

As we see we can arrive at the solutions much faster than Balakrishnan and Wong [1], though in interaction of the human decision maker. Because each step is very fast, the overall process of constructing an optimal solution still does not take very long.

**Problem 5**: This problem is a larger problem first reported in [5]. Characteristics of this problem are:
Number of employees is 17 (length of planning period is 17 weeks).

Table 7: First Class Schedule solution of problem from [7]

| Employee/day | Mon | Tue | Wed | Thu | Fri | Sat | Sun |
|---|---|---|---|---|---|---|---|
| 1 | D | D | D | D | D |   |   |
| 2 |   | D | D | D | D | D | D |
| 3 | D |   |   | N | N | N | N |
| 4 |   |   |   |   | A | A | A |
| 5 | A | A | A | A |   |   |   |
| 6 | N | N | N | N | N |   |   |
| 7 |   |   | A | A | A | A | A |
| 8 | A | A |   |   |   | N | N |
| 9 | N | N | N |   |   | D | D |

Three nonoverlapping shifts.
Temporal requirements are:

$$R_{3,7} = \begin{pmatrix} 5 & 4 & 4 & 4 & 4 & 4 & 3 \\ 5 & 4 & 4 & 4 & 4 & 4 & 4 \\ 4 & 3 & 3 & 3 & 4 & 4 & 4 \end{pmatrix}$$

Constraints:

- Rest-period lengths must be between 2 and 7 days

- Work-periods lengths must be between 3 and 8 days

- A shift cannot be assigned to more than 4 consecutive weeks in a row

- Shift changes are allowed only after a rest period that includes a Sunday or a Monday or both

- The only allowable shift changes are 1 to 3, 2 to 1, and 3 to 2

Balakrishnan and Wong [1] need 457.98 seconds to arrive at the optimal solution which is given in Table 8.

With First Class Schedule we cannot model constraints 3, 4, and 5 in their original form. We allow changes in the same block and for these reason we have other shift change constraints. In our case the following shift changes are not allowed: 2 to 1, 3 to 1, and 3 to 2. Additionally, we limit the rest period length from 2 to 4 and work periods length from 4 to 7. Maximum and minimum length of blocks of successive shifts are given with vectors $MAXS_3 = (7, 6, 5)$ and $MINS_3 = (2, 2, 2)$.

With these conditions the first class solution {6 6 5 5 5 5 5 5 5 5 5 5 5 5 5} is found after 0.22 and the first 9 solutions after 14.2 seconds. Of course there exist much more class solutions, but finding all class solutions will take too much time for this large problem. If we choose the first solution with the most optimal blocks we will obtain solutions with 5 weekends off, even though the weekends are one after each other. We arrive at a better solution with the following class solution: {7 6 5 5 5 5 5 5 5 5 5 5 5 5 4}. In step 2 we stop the process of generation of distributions of work and days-off blocks after 20 seconds and we get 3 solutions. From these solutions we select the solution with the order of blocks (7 6 5 5 5 5 5 5 5 5 5 5 5 4 5 5 ). The first solution (steps 3 and 4) is generated after 0.73 seconds and the first 50 solutions after 4.67 seconds. The first solution is given in Table 9.

Table 8: Solution of Balakrishnan and Wong [1] to the problem from [5]

| Employee/day | Mon | Tue | Wed | Thu | Fri | Sat | Sun |
|---|---|---|---|---|---|---|---|
| 1  | D | D | D | D | D | D | D |
| 2  | D |   |   | D | D | D | D |
| 3  | D | D | D |   |   |   |   |
| 4  | D | D | D | D | D | D |   |
| 5  |   |   |   | N | N | N | N |
| 6  | N | N |   |   | N | N | N |
| 7  | N | N | N |   |   |   |   |
| 8  | A | A | A | A | A | A | A |
| 9  | A |   |   | A | A | A | A |
| 10 | A | A | A |   |   |   |   |
| 11 | D | D | D | D | D | D | D |
| 12 |   |   | N | N | N | N | N |
| 13 | N |   |   |   | N | N | N |
| 14 | N | N | N | N |   |   |   |
| 15 |   | A | A | A | A | A | A |
| 16 | A |   |   | A | A | A | A |
| 17 | A | A | A |   |   |   |   |

As you can see the solution has a much worse distribution of weekends than the solution from [1] but our solution has no blocks of length 8 and has many optimal blocks (of length 5). Our solutions also have not more than 5 successive night shifts (seven night shifts are considered too much).

Much better distribution of weekends-off can be found with FCS if the maximum length of work days is increased to 8. In this case step 2 of FCS takes longer because it depends directly on the number of blocks.

One disadvantage of FCS is that the user has to try many class solutions to find an optimal solution. However, the time to generate solutions in each step is so short that interactive use is possible. Other advantages of interactively solving these scheduling problems is the possibility to include the user in the decision process. For example one may prefer longer blocks but better

Table 9: First Class Schedule solution to the problem from [5]

| Employee/day | Mon | Tue | Wed | Thu | Fri | Sat | Sun |
|---|---|---|---|---|---|---|---|
| 1 | D | D | D | D | D | D | D |
| 2 |   |   | D | D | D | D | D |
| 3 | D |   |   | D | D | D | D |
| 4 | D |   |   | D | D | D | A |
| 5 | A |   |   | A | A | A | A |
| 6 | A |   |   | A | A | A | N |
| 7 | N |   |   |   | A | A | A |
| 8 | A | A |   |   |   | A | A |
| 9 | A | A | A |   |   | N | N |
| 10 | N | N | N |   |   | N | N |
| 11 | N | N | N |   |   | N | N |
| 12 | N | N | N |   |   |   |   |
| 13 | D | D | D | N | N |   |   |
| 14 |   | A | A | A | N | N |   |
| 15 |   | D | D | N | N |   |   |
| 16 | D | D | A | A | A |   |   |
| 17 | A | A | A | N | N |   |   |

distribution of weekends to shorter work blocks but worse distribution of weekends.

## 5 Conclusions

In this paper we proposed a new framework for solving the rotating workforce scheduling problem. We showed that this framework is very powerful for solving real problems. The main features of this framework are the possibility to generate high quality schedules through the interaction with the human decision maker and to solve real cases in a reasonable amount of time. Besides making sure that the generated schedules fulfill all hard constraints, it also allows to incorporate preferences of the human decision maker regarding soft constraints that are otherwise more difficult to assess and to model. In step 1 an enhanced view of possible solutions subject to the length of work blocks is given. In step 2 preferred sequences of work blocks in connection with weekends off features can be selected. In step 4 bounds for successive shifts and shift change constraints can be specified with much more precision because the decision maker has a complete view on terms (shift sequences) that are used to build the schedules. Step 2 of our framework can be solved very efficiently because the search space has already been much reduced in step 1. Furthermore, in step 4 we showed that assignment of shifts to the employees can be done very efficiently with backtracking algorithms even for large instances if sequences of shifts for work blocks are generated first. When the number of employees is very large they can be grouped into teams and thus again this framework can be used. Even though this framework is appropriate for most real cases, for large instances of problems optimal solution for weekends off cannot always be guaranteed because of the size of the search space. One possibility to solve this problem more efficiently could be to stop backtracking when one solution that has the or almost the maximum number of weekends off is found (for a given problem we always know the maximum number of weekends off from the temporal requirements). Once we have a solution with most weekends off, other search technique like local search can be used to improve the distribution of weekends off. Finally this framework can be extended by introducing new constraints.